\documentclass{emulateapj}
\usepackage{natbib}
\usepackage{epsfig}
\newcommand{\KMS}{\mbox{km s}^{-1}}

\def\ltsima{$\; \buildrel < \over \sim \;$}
\def\ltsim{\lower.5ex\hbox{\ltsima}}
\def\gtsima{$\; \buildrel > \over \sim \;$}
\def\gtsim{\lower.5ex\hbox{\gtsima}}

\def\Msun{\>{\rm M_{\odot}}}

\def\MPR#1{{\it Moving Puncture Recipe}#1 (MPR#1)\gdef\MPR{MPR}}
\def\ahz#1{apparent horizon#1 (AH#1)\gdef\ahz{AH}}
\def\CLA#1{close-limit approximation#1 (CLA#1)\gdef\CLA{CLA}}
\def\pnw#1{post-Newtonian#1 (PN#1)\gdef\pnw{PN}}
\def\qnm#1{quasi-normal mode#1 (QNM#1)\gdef\qnm{QNM}}
\def\isco#1{innermost stable circular orbit#1 (ISCO#1)\gdef\isco{ISCO}}
\def\eos#1{equation of state#1 (EOS#1)\gdef\eos{EOS}}
\def\ns#1{neutron star#1 (NS#1)\gdef\ns{NS}}
\def\bbh#1{binary black holes#1 (BBH#1)\gdef\bbh{BBH}}
\def\bhns#1{black hole -- neutron star#1 (BHNS#1)\gdef\bhns{BHNS}}
\def\nsns#1{neutron star -- neutron star#1 (NSNS#1)\gdef\nsns{NSNS}}
\def\emri#1{extreme mass-ratio inspiral#1 (EMRI#1)\gdef\emri{EMRI}}
\def\emrb#1{extreme mass-ratio binaries#1 (EMRB#1)\gdef\emrb{EMRB}} 
\def\grb#1{gamma-ray burst#1 (GRB#1)\gdef\grb{GRB}}
\def\imbh#1{intermediate mass black hole#1 (IMBH#1)\gdef\imbh{IMBH}}
\def\smbh#1{supermassive black hole#1 (SMBH#1)\gdef\smbh{SMBH}}
\def\bh#1{black hole#1 (BH#1)\gdef\bh{BH}}
\def\ulx#1{ultra-luminous x-ray source#1 (ULX#1)\gdef\ulx{ULX}}
\def\lmxbs{low-mass x-ray Binaries (LMXBs)\gdef\lmxbs{LMXBs}\gdef\lmxb{LMXB}} 
\def\lmxb{low-mass x-ray Binary (LMXB)\gdef\lmxbs{LMXBs}\gdef\lmxb{LMXB}} 

\begin{document}
\title{Gravitational Wave Recoil and the Retention of Intermediate Mass Black Holes}

\author{Kelly Holley-Bockelmann\altaffilmark{1}, Kayhan G\"{u}ltekin\altaffilmark{2}, Deirdre Shoemaker\altaffilmark{1}, Nico Yunes\altaffilmark{1}}

\altaffiltext{1}{IGPG, Center for Gravitational Wave Physics,
The Pennsylvania State University, University Park, PA 16802, USA
Email: kellyhb@gravity.psu.edu, deirdre@gravity.psu.edu, yunes@gravity.psu.edu}

\altaffiltext{2}{Department of Astronomy,
University of Michigan, Ann Arbor, MI, 48109, USA
Email:kayhan@umich.edu}

\begin{abstract}
  During the inspiral and merger of a binary black hole, gravitational
  radiation is emitted anisotropically due to asymmetries in the
  merger configuration. This anisotropic radiation leads to a
  gravitational wave kick, or recoil velocity, as large as $\sim 4000
  \KMS$.  We investigate the effect gravitational recoil has on the
  retention of intermediate mass black holes (IMBH) within Galactic globular
  clusters. Assuming that our current understanding of IMBH-formation
  is correct and yields an IMBH-seed in every globular cluster, we
  find a significant problem retaining low mass IMBHs ($\ltsim 1000~\Msun$) 
  in the typical merger-rich globular cluster environment. Given a uniform 
  black hole spin distribution and orientation and a Kroupa IMF, we find that 
  at most $3\%$ of the globular clusters can retain an IMBH larger than
  $1000 M_\odot$ today.  For a population of black holes that better 
  approximates mass loss from winds and supernovae, we find 
  that $16\%$ of globulars can retain an IMBH larger than $1000~\Msun$.
  Our calculations show that if there are black holes of mass $M > 60~\Msun$ 
  in a cluster, repeated IMBH-BH encounters will eventually eject 
  a $M = 1000~\Msun$ IMBH with greater than 30\% probability.
  As a consequence, a large population
  of rogue black holes may exist in our Milky Way halo. We discuss the 
  dynamical implications of this subpopulation, and its possible connection to
  ultraluminous X-ray sources (ULXs).

\end{abstract}


\section{Introduction}\label{sec:intro}

Ample observational evidence exists for two types of black holes:
stellar mass (BH), with~$10 \Msun \ltsim m \ltsim 10^2 \,\Msun$, and
supermassive (SMBH), with $m \gtsim 10^6 \,\Msun$ (e.g.,
\cite{kormendy1995}).  Although the existence of a third black hole
class is still under debate, there are observational hints for
intermediate mass black holes (IMBHs) as well, with
masses~$10^2\,\Msun \ltsim m \ltsim 10^5 \,\Msun$ (e.g., 
\cite{Gebhardt05:imbh,Gerssen02:imbh, Filippenko03:imbh,2006MNRAS.368..677H,2007MNRAS.374..344T,2007MNRAS.374..857T,2006astro.ph.12040T,2007ApJ...661L.151U}; 
c.f. \cite{Baumgardt:03noimbh} for an alternative view.) These IMBHs may
form within dense star clusters and may be the best explanation for
ultraluminous X-ray sources (ULXs) in young star forming regions
\citep{Fabbiano:89xray,Roberts:00xray,Ptak:04xray,Fabbiano:06xray} and
in nearby extragalactic star clusters.

If the observational case for IMBHs is still unclear, the exact
formation mechanism is perhaps less constrained. Within a globular
cluster, there are currently three plausible IMBH formation theories:
stellar runaway; compact object mergers; and growth from Population
(Pop) III remnants (cf. \citet{vandermarel:2004ib} for a review.)
Each of these mechanisms starts with very different initial conditions, and 
the first two require the high stellar densities of a globular or stellar 
cluster environment. Thus, growing IMBHs over the ensemble of observed
globular clusters may require more than one single mechanism. We
briefly introduce these mechanisms below.
 
Recent simulations of core collapse in stellar clusters have shown
that stellar collisions could induce rapid {\it stellar runaway}
growth of an ${\cal{O}}(10^3 \Msun)$ IMBH in as little as 3 Myr
\citep{Portegies:2002rg,Gurkan:2005mb,Freitag:2005rc}; see also
\citep{Portegies:2004fm}. However, this runaway growth process
requires number densities higher than ${\cal{O}}(10^6
/{\mathrm{pc}}^3)$ and few (if any) globular clusters in the Milky Way
are this dense today. Of course, globular clusters are thought to be
more dense at formation \citep{Trenti:07imbh,Heggie:06imbh}, but
calculations still suggest that the runaway process may only occur in
about $20\%$ of the Milky Way globular cluster system
\citep{Baumgardt:05imbhgc}. By using less extreme central densities
and including the effect of primordial binaries, simulations have
shown that instead of a single IMBH, stellar runaways 
generate two or more black holes with masses $\sim 500 M_\odot$
\citep{Gurkan:2005mb}. When these large black holes merge they form a single
IMBH and a strong gravitational wave signal for the Laser Interferometer Space Antenna, or LISA, a planned gravitational wave observatory set to launch in the
next decade\citep{Fregeau:06imbhgw,Gurkan:2005mb}.

While runaway star collisions seem to be the preferred IMBH formation channel
within dense systems, in systems where the relaxation time is long, {\it stellar mass BH collisions} may
be a more likely growth
mechanism. Here the initial seed IMBH forms naturally
from a heavier than average black hole, perhaps a supernova remnant
with $\sim 250 \Msun$ \citep{Wise:05snpop3}. Then, stellar mass BHs
are captured by the initial seed, and the resulting binary hardens through
interactions with other stellar-mass objects. Although these interactions risk
ejecting the BH supply before significant growth occurs, one way to
funnel more BHs to the growing IMBH is through 4-body collisions
\citep{Oleary:2005bm,Miller:2002pi,Gultekin:2004gi}.

Significantly higher mass {\it Pop~III stars} may yield IMBHs
directly. If a Pop~III protostar has a mass of $~10^5 M_\odot$, it
will be gravitationally unstable and will collapse to an IMBH before
it even enters the main sequence \citep{Baumgarte99e,Shibata02}. The
probability of an IMBH forming from such a massive Pop~III star
certainly depends on the initial stellar mass function (IMF), as well
as highly uncertain details of zero-metallicity stellar evolution.
However, there has been some suggestions that the IMF in the early
universe is quite top-heavy
\citep{Schneider:02topheavy,Abel:02firststar}, and that stellar mass
loss is negligible \citep{Fryer:01spin,Heger:03sn}.  These suggestions
imply that BHs in a proto globular cluster
environment are more massive than those formed in more recent times
\citep{vandermarel:2004ib}. On the other hand, Pop~III star formation
is thought to take place in dark matter overdensities at $z\sim 12-20$
\citep{Madau:01Pop3} and only the most massive globular clusters are
thought to be embedded in a significant enough dark matter overdensity
to allow for Pop~III formation.

However the IMBH forms, for the first $\sim 0.5~{\mathrm{Gyr}}$ after
formation it is dynamically active \citep{Spitzer:1987gcbook}. Due to
mass segregation, the initial environment around an IMBH in a globular
cluster is especially rich in BHs \citep{Fregeau:02msgc}. While many
of these BHs are quickly ejected by few body interactions with the
IMBH, enough remain to subject the IMBH to $10~{\mathrm{s}}-100~{\mathrm{s}}$ 
mergers with BHs in the primordial
globular cluster system
\citep{Portegies:00bhmerge,Oleary:2005bm,Gultekin:2004gi}.

In light of general relativistic black hole merger simulations,
surviving this IMBH-BH merger epoch may be difficult. Recent advances
in numerical relativity have at last pinned down the dynamics of black
hole mergers -- simulating the coalescence, merger, and ringdown of
equal-mass circular non-spinning binary black holes in full general
relativity ~\citep{Pretorius:2005gq,Campanelli05a,Baker:2006yw}. In
the past year, more astrophysically relevant numerical simulations,
including spins and unequal masses, have been published by several
groups.

One of the most exciting results of general relativity for structure
formation is that binary black hole systems strongly radiate linear
momentum in the form of gravitational waves during the plunge
phase of the inspiral. This radiation directly results from an
asymmetry in the orbital configuration and can generically yield a
gravitational wave ``kick'' velocity as fast as $4000~\KMS$
~\citep{Gonzalez:2007hi,Campanelli:2007cg}. Even typical kick
velocities ($\sim 200~\KMS$) are interestingly large when compared to
the escape velocity of an average globular cluster ($\sim 50 \KMS$)
\citep{Webbink:85vesc,Favata:2004wz,Merritt:2004xa}.  Hence, regardless of how
  an IMBH forms, the biggest challenge may be to determine  how a
  globular cluster {\it retains} them in the face of a repeated onslaught
of gravitational wave kicks from mergers with other black holes.  
Of course if IMBHs are formed through mergers of stellar-mass black
holes, the formation mechanism itself needs to be able to account for
these large kicks as well.

In this paper we explore IMBH retention within young globular clusters
after collisions with BHs.  We calculate the retention probability
under a variety of assumptions for the initial IMBH seed mass, the BH
mass distribution, and the initial spin distributions.  In addition, for
the mass range encompassed by the IMBH formation channels listed above (stellar runaway,
stellar-mass BH collisions, Pop III stars), we simulate the IMBH
survival probability after merging with the available BHs over a
collision timescale. Given the known current structure of the Milky
Way globular cluster system, we can then estimate the fraction of
globular clusters that may have allowed a particular IMBH formation
channel. Combining these two key estimates, we can determine the
maximum expected number of Milky Way globular clusters that could have
retained their IMBHs. We outline the approach in Sec.~2, present the
results in Sec.~3 and discuss the caveats, implications, and future
directions in Sec.~4.

\section{Assigning Kicks}

Gravitational recoil estimates of binary black hole mergers have been
addressed using both semi-analytic methods in the unequal-mass,
non-spinning
case~\citep{Fitchett:1983fc,Favata:2004wz,Damour:2006tr,Blanchet:2005rj,2006PhRvD..74l4010S}
and numerical methods in more general, spinning, unequal mass
scenarios~\citep{Herrmann:2006ks,Baker:2006vn,Gonzalez:2006md,Herrmann07,Koppitz:07kick,gonzalez-2007,campanelli-2007-659,campanelli-2007,tichy-2007}.
For non-spinning binaries, the most comprehensive numerical study of
kicks is that of~\citet{Gonzalez:2006md}, which was found in agreement
with the semi-analytic estimates of~\citep{2006PhRvD..74l4010S}, where
a maximum kick velocity of $\sim 175\ \pm 10 \; \KMS$ was obtained for
mass ratio $q=M_1/M_2 \sim 0.36 \pm 0.02$).

\begin{figure*}
\begin{center}
  \epsfig{file=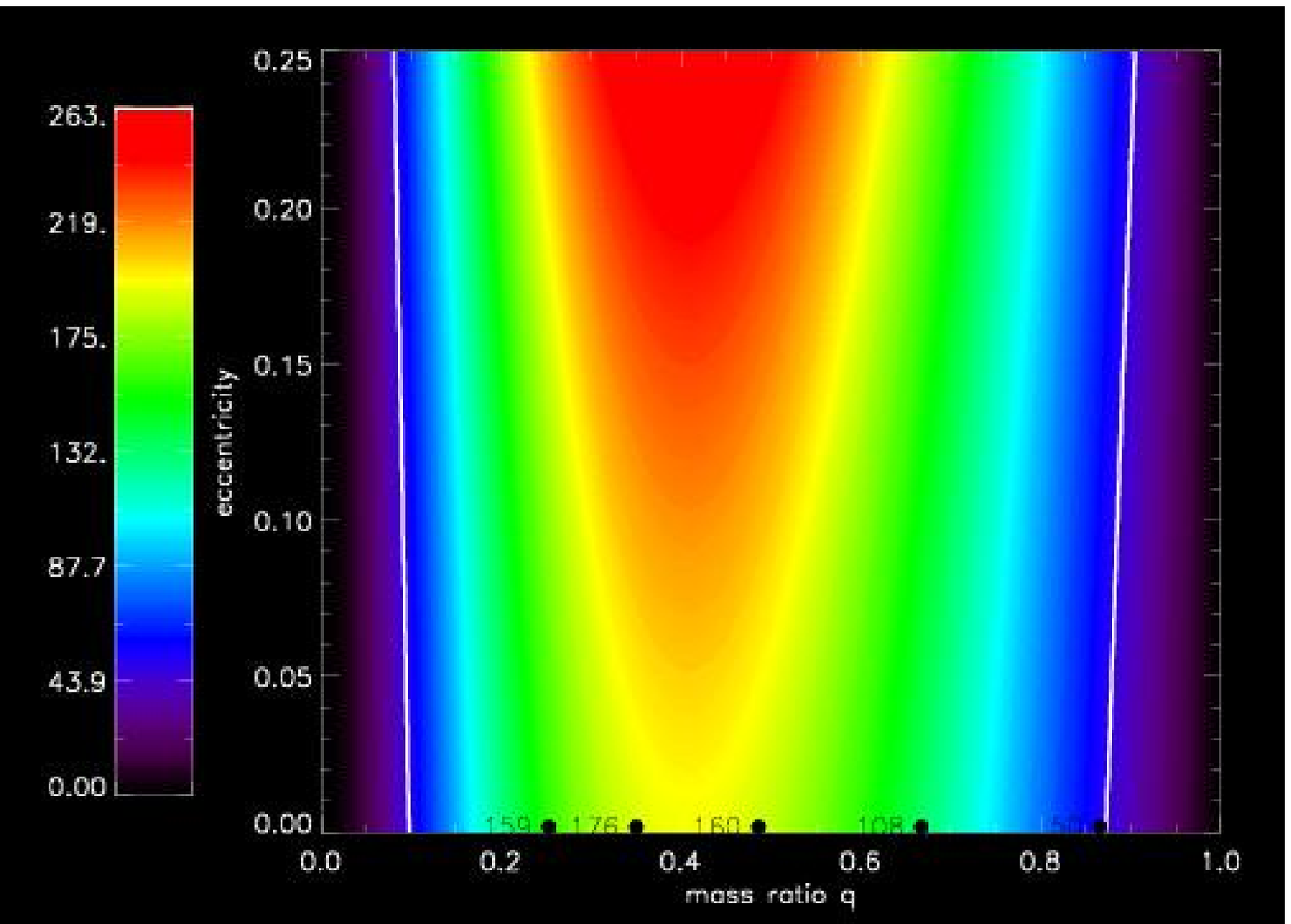, height=2.25in, width=3.15
    in}\space\space\space\space\space\space\space\space\space\space\epsfig{file=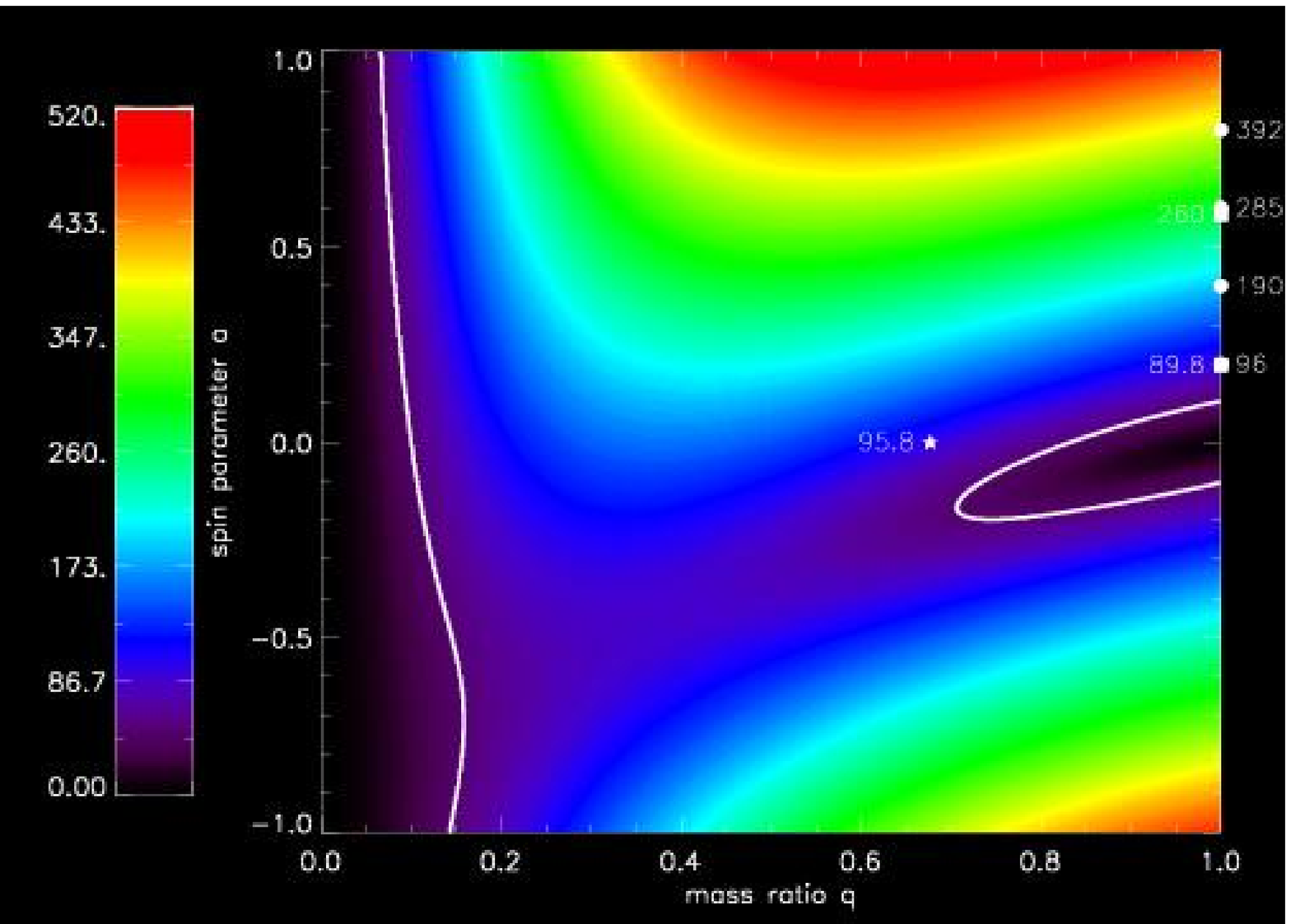,
    height=2.25in, width=2.95in}
\caption{
  Left: The gravitational wave kick velocity from a non-spinning black
  hole merger as a function of mass ratio and eccentricity. The white
  contour marks $50~\KMS$, the escape velocity of a typical globular
  cluster. Overlaid are data from numerical relativity merger
  simulations~\citep{Gonzalez:2006md} (black dots on zero-eccentricity
  line.) Note that most of this plane has not been confirmed by direct
  numerical experiment, but has been extrapolated from semi-analytic
  studies~\citep{2006astro.ph.11110S}. In this plane, only mergers
  with the largest mass ratios will remain in the globular cluster,
  which argues for a very high-mass initial seed.  Right: The
  gravitational wave kick velocity from spinning black hole mergers on
  a circular orbit as a function of spin magnitude and mass ratio.
  Here, the spins of the black hole are anti-aligned to one another
  and are perpendicular to the orbital plane of the binary black hole
  before merger. Note that other spin orientations yield even higher
  kicks.  As in the previous figure, the $50~\KMS$ contour is
  overplotted, as well as selected results from full numerical
  relativity simulations: squares are from \citet{Koppitz:07kick};
  circles are from ~\citet{Herrmann07}, and star is from
  ~\citet{Bruegmann03}. This figure suggests
  that globular clusters can retain IMBHs only when they merge with
  other BHs with low spin and high mass ratio.}
\label{Fig-v-recoil}
\end{center}
\end{figure*}

The gravitational recoil is expected to increase with increasing spin~\citep{Redmount:1989rr,WhitbeckPhD}, and this behavior has been confirmed by several numerical relativity groups. Numerical simulations
have demonstrated that the radiative linear momentum
loss predicted by Post-Newtonian studies~\citep{RevModPhys.52.299,PhysRevD.52.821} can be used to fit
numerical results, yielding a generalized formula for the recoil
velocity as a function of the individual black hole's spin, initial
orientation, phase at merger and mass ratio. We here adopt the
parameterized fit of~\citet{campanelli-2007-659} adding the expected
$(1+e)$ contribution for eccentric orbits~\citep{2006astro.ph.11110S}
to yield the following formula:
\begin{equation}
v_{\mathrm{kick}} = \left(1+e\right) \left[{\hat{x}} \; \left(v_{m} +
  v_{\perp} \cos{\xi}\right)  + {\hat{y}} \; v_{\perp} \sin{\xi}  + 
  {\hat{z}} \; v_{\parallel}\right],  
\label{eqn:Fit}
\end{equation}
where
\begin{equation}
v_{m} = A \frac{q^2 \left(1-q\right)}{\left(1+q\right)^5} \left[ 1+
  B \frac{q}{\left(1+q\right)^2}\right], 
\label{eqn:Fit1}
\end{equation}
\begin{equation}
v_{\perp} = H \frac{q^2}{\left(1+q\right)^5} \left( \alpha_2^\parallel
  - q \alpha_1^\parallel\right), 
\label{eqn:Fit2}
\end{equation}
and
\begin{equation}
v_{\parallel} = K \cos\left(\Theta-\Theta_0\right) \frac{q^2}{\left(1+q\right)^5}
\left( \alpha_2^\perp - q \alpha_1^\perp \right),
\label{eqn:Fit3}
\end{equation}
where the fitting constants are~$A=1.2 \times 10^{4} \; \KMS$, $B =
-0.93$, $H=(7.3 \pm 0.3) \times 10^{3} \; \KMS$, and $K=(6.0 \pm 0.1)
\times 10^{4} \; \KMS$, while the subscripts $1$ and $2$ refer to the
first and second BH respectively. The unit vectors $(\hat{x},\hat{y})$
are orthogonal to each other and span the initial orbital plane, while
$\perp$ and $\parallel$ stands for perpendicular and parallel to the
orbital angular momentum.  There are $4$ fitting parameters: the mass ratio
$q\equiv M_2/M_1$; the reduced spin parameter $\alpha_i \equiv
S_i/M_i^2$, where $S_i$ is the spin angular momentum of BH $i$; and
the eccentricity $e$.  In addition, there are 3 angles to specify the
orientation of the merger: $\Theta$, the angle between the
``in-plane'' component of $\delta^i \equiv (M_1 + M_2) \left(S_2^i/M_2
  - S_1^i/M_1\right)$ and the infall direction at merger; $\Theta_0$,
the angle between $\delta^i$ and the initial direction of motion; and
$\xi$, the angle between the unequal mass and spin contribution to the
recoil in the orbital plane.  The merger phase within the orbital
plane may also play a role, but it is not included explicitly in this
fit. The recoil velocities as given in Eq.~(\ref{eqn:Fit3})
are plotted in Fig.~\ref{Fig-v-recoil} as a function of mass ratio and
spin parameter.

In order to compute the probability that an IMBH remains in its
globular cluster, our calculation shall proceed as follows. First we
begin by assuming that an IMBH has formed within a globular cluster
with a particular initial mass, $M_{\mathrm{IMBH}}$, which we shall
vary.  Second, we further assume a certain mass distribution for the
BHs in the vicinity of the IMBH, which we shall also vary. Third, we
subject this IMBH to a number of mergers expected within a
proto globular cluster environment that we shall describe below.
Finally, we determine the probability that the kick velocity for the
IMBH has remained below the canonical globular cluster escape velocity
($50 \KMS$) during the entire chain of mergers.

Even in the absence of an IMBH, BHs eject themselves from globular
clusters via standard few-body interactions on a timescale of $\sim 1~{\rm Gyr}$
after the onset of mass segregation
\citep{Kulkarni:93bhgc,Sigurdsson:93bhgc,Portegies:00bhmerge,Oleary:2005bm}.
Therefore, due to such Newtonian few-body interactions, the supply of
BHs is eventually depleted.  With an IMBH, however, this process
speeds up impressively, as ejections by interactions with an IMBH
become the dominant source of stellar-mass BH ejections
\citep{Gultekin:2006tb}.  As in most few-body interactions, the
ejection of one object tightens the orbit of a remaining bound pair,
in this case an IMBH-BH binary -- and after several subsequent
ejections, the IMBH-BH binary merges.  Soon after all the BHs have
been evacuated, the short epoch of IMBH-BH mergers ends.

\begin{figure*}
\begin{center}
  \epsfig{file=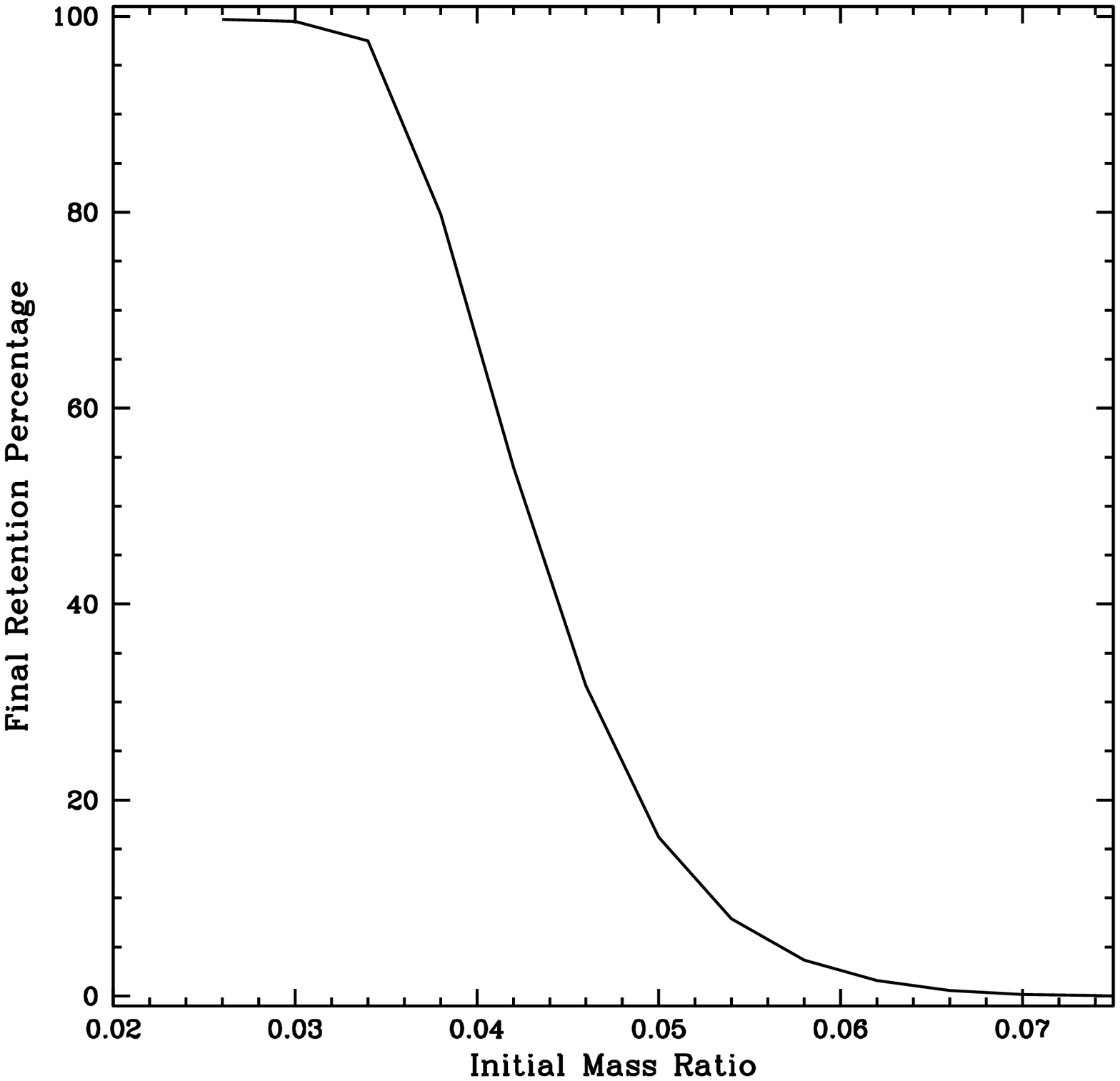, height=2.in, width=2.
    in}\space\space\space\space\space\space\space\space\space\space\epsfig{file=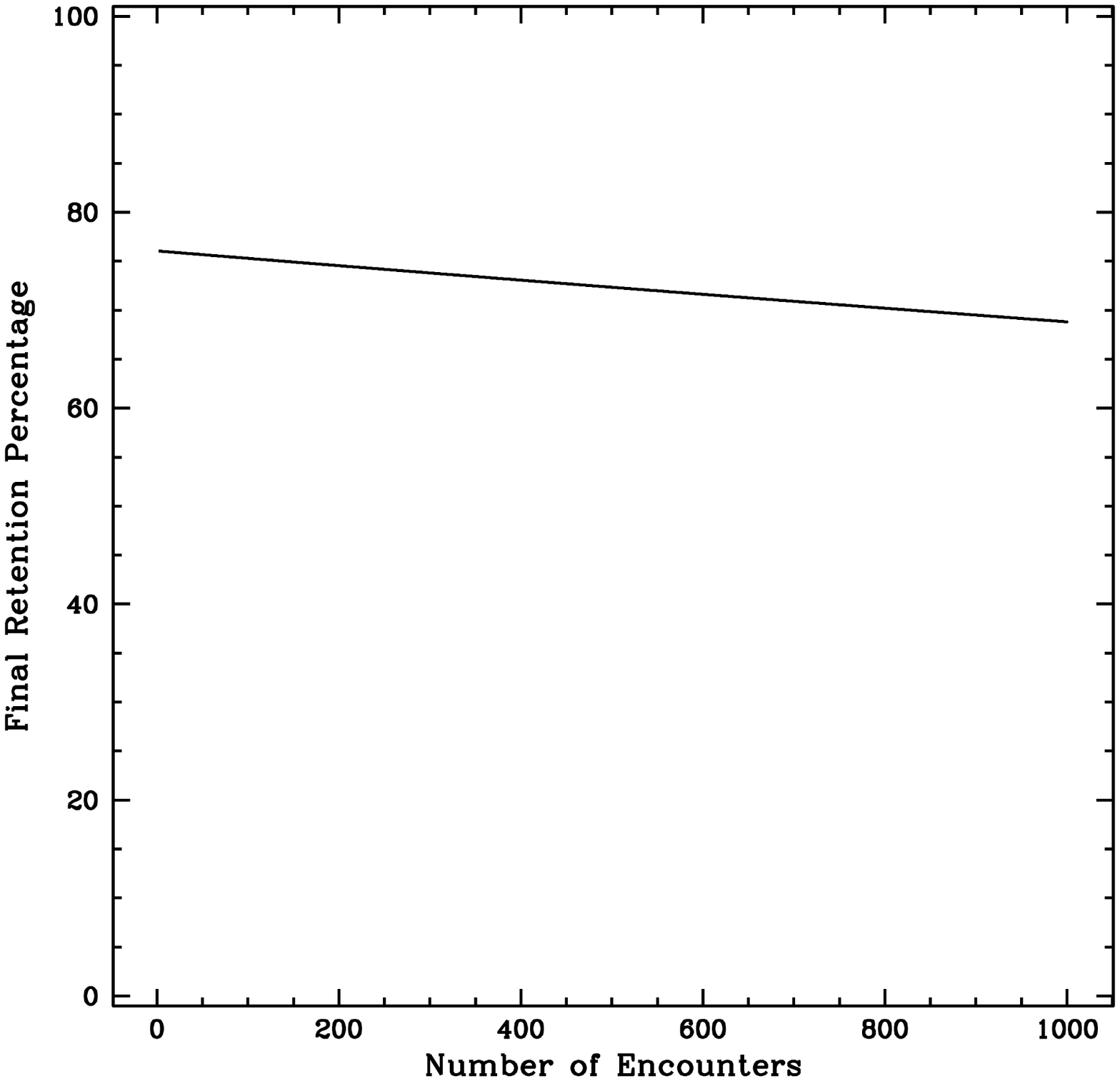,
    height=2.in,
    width=2.in}\space\space\space\space\space\space\space\space\space\space\epsfig{file=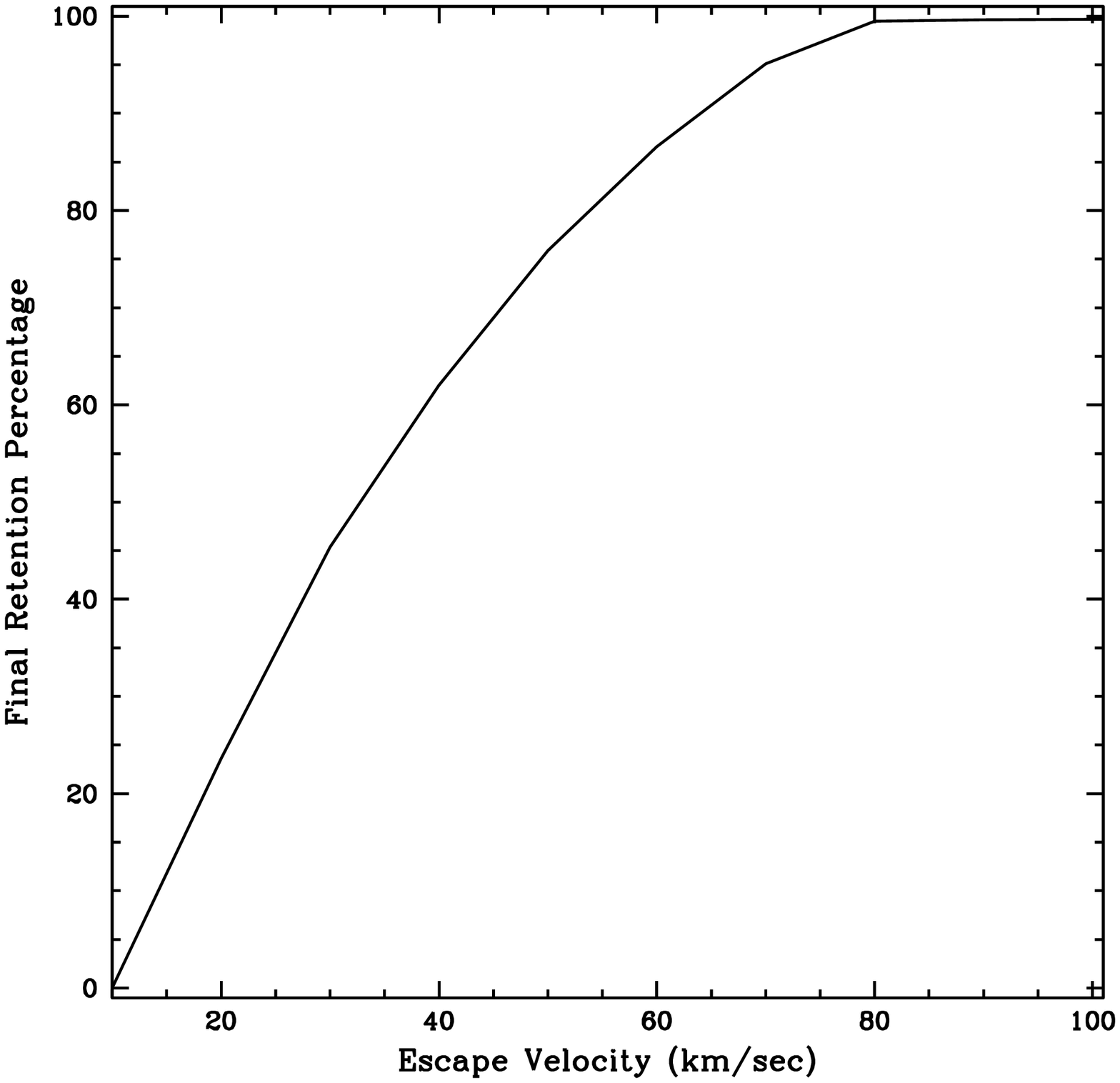,
    height=2.in, width=2.in}
\caption{
  Left: Retention probability as a function of mass ratio of the
  interacting black holes. We assume an initial IMBH mass of
  $M_{\mathrm{IMBH}}=500~M_\odot$, an escape velocity of $50~\KMS$,
  and a merger chain of $25$ interactions. Mergers are chosen with
  random orientations and spin magnitudes.  Middle: Retention
  probability as a function of the number of BH mergers. We assume
  $M_{\mathrm{sec}}=20~M_\odot$, $M_{\mathrm{IMBH}}=500~M_\odot$, and
  random orientations and spin magnitudes. Right: Retention
  probability as a function of the escape velocity of the host
  structure. We assume $M_{\mathrm{sec}}=20~M_\odot$,
  $M_{\mathrm{IMBH}}=500~M_\odot$, and random orientations and spin
  magnitudes.}
\label{Fig-ret-prob} 
\end{center}
\end{figure*}

Within this theoretical framework, it is possible to construct a
fiducial number of mergers for a proto globular cluster. This number
can be written as
\begin{equation}
N_{\mathrm{merge}} = \frac{N_{\mathrm{BHs}}}{n_{\mathrm{ejections/merger}}}.
\label{eqn:NN}
\end{equation}
Gultekin et al. 2006 predict $N_{\mathrm{merge}} \sim 25$ per IMBH,
and we adopt this for the fiducial number of mergers that the IMBH
encounters. Although we do vary this parameter in
figure~\ref{Fig-ret-prob}, the dependence of the retention probability
on the number of mergers is relatively minor, since the IMBH grows in
mass over each merger and the kick velocity increases with increasing
mass ratio.
 
In order to assign a kick velocity to each of the mergers, we choose
the orientation, spin, mass, and eccentricity. We outline the
assumptions made for each distribution below. Let us first discuss the
issue of the initial spin orientation. Hydrodynamic interactions
between a gas disk and a black hole binary are believed to align the
spin directions to the angular momentum axis of the binary orbital
plane in many active galaxies~\citep{tamara:07spin}. However, the
environment of a globular cluster is not particularly gas-rich, so
there is no {\it{ab initio}} reason to expect the black hole spins to
be aligned. We therefore assume an isotropic distribution of
orientation angles for each encounter.

Let us now discuss the choice of spin magnitude. Most theories predict
a non-zero spin for a black hole produced via stellar runaway
\citep{Rees:07mbh} or from a supernovae remnant \citep{Fryer:01spin}.
If an IMBH started with zero spin, a merger is likely to spin up the
remnant through transfer of orbital to spin angular
momentum~\citep{Gammie:2003qi}.  However, a Kerr black hole can $spin
 down$ when magnetic field lines thread through the ergosphere
to magnetically brake the system \citep{Blandford:77spin}, provided
there is a gaseous disk around the remnant. Taking all these
consideration into account, we shall explore three cases: (1) the spin
magnitude is selected from a uniform initial spin distribution
(fiducial case); (2) the initial spin magnitude of the IMBH seed is set to
$0.998 M_{\mathrm{IMBH}}^2$; and (3) the spin is initially set to zero. We
assume the spin of the secondary BH to be randomly selected from a
distribution of $[0,0.998]~M_{\mathrm{sec}}^2$, where
$M_{\mathrm{sec}}$ is the mass of the secondary BH. Since
these stellar-mass BHs originate as a supernova remnant, though, the spin is
likely to be high \citep{Heger:03sn}. Therefore, since the kick
velocity increases with increasing spin, this shall lead to a
conservative survival probability for each merger tree.

The survival probability shall be studied as a function of the initial
mass of the seed IMBH. We choose a range of $10-3000 M_\odot$ to
encompass the plausible IMBH formation channels and relevant masses 
involved. For the three formation channels discussed in \S~\ref{sec:intro}, 
the seed masses are likely to be the following: $\sim 1000
M_\odot$ for a single stellar runaway~\citep{Portegies:2004fm}; $\sim
100 M_\odot$ for the stellar mass BH collision channel, where the seed
IMBH is produced by a massive supernova remnant or a small stellar 
runaway~\citep{Heger:03sn}; and $\sim 200-400 M_\odot$ for Pop~III
remnants within a dense globular cluster~\citep{Wise:05snpop3,Madau:01Pop3}. 

One of the biggest uncertainties is determining a proper distribution
for the secondary BH masses.  As figure~\ref{Fig-ret-prob} shows, the
retention probability depends strongly on the mass ratio between the
IMBH and BH. Theoretical black hole mass 
distributions from solar metallicity field populations tend to peak 
strongly around $10 M_\odot$~\citep{2001ApJ...554..548F}. 
There are, however, several strong competing effects in a primordial globular 
cluster that can change this distribution (e.g., low metallicity stellar evolution, 
high binary fraction, stellar collisions, mass segregation, and
natal kicks). 
 For our fiducial experiment, we assume that the
secondary masses are selected from a Kroupa IMF with an upper mass
cutoff of $120 M_\odot$ \citep{Kroupa:01imf}. Recall that this yields
an average stellar mass of about $1 M_\odot$, much smaller than the suspected
average stellar mass from a zero-metallicity environment.  We further
assume that each star above $10 M_\odot$ evolves directly to a BH with
no mass loss.  This simplified treatment 
gives us an average BH mass of $\sim 20 M_\odot$. Clearly, mass loss
would decrease the average BH mass even in these low metallicity primordial 
globulars.  Therefore, we used a more sophisticated black hole mass function 
that approximates the results of model C1 of \citet{Belczynski:2006bh}.  This model includes the
effect of mass loss from supernovae and winds for a population of stars with metallicity $Z = 0.0001$, though it has fewer massive remnants from binary mergers (cf Figure 8  of \citet{Belczynski:2006bh}) than is expected for a primordial globular cluster.

\begin{figure}
  \epsfig{file=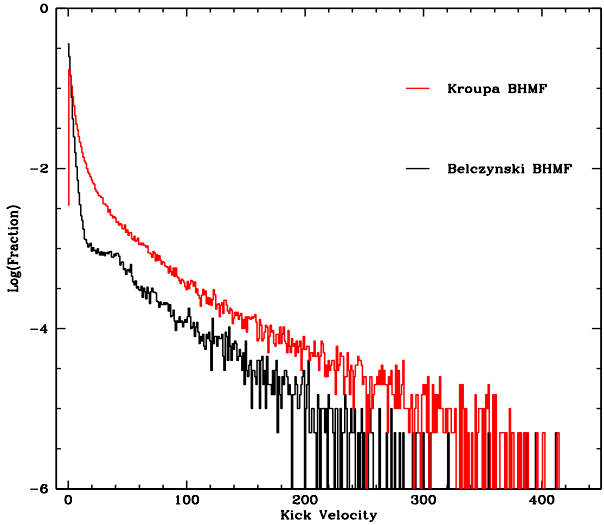, height=3.in, width=3.4in}
\caption{Fraction of encounters as a function of kick velocity for two
black hole mass functions. We assume the IMBH has a mass of $1000 M_\odot$
and that the spin magnitudes and orientations are distributed uniformly.}
\label{fig:vkickbhmf} 
\end{figure}

Figure~\ref{fig:vkickbhmf} demonstrates the effect on the kick velocity distribution between
these two black hole mass functions for a $1000 M_\odot$ IMBH.
Since the distribution of secondary 
masses is so uncertain, we demonstrated the effect of varying the 
mass ratio in Figure~\ref{Fig-ret-prob}.  
For a near-solar metallicity stellar cluster, the entire population of
black holes may be less than $M \ltsim 20~M_\odot$
\citep{2001ApJ...554..548F}.

Finally, we must select the eccentricity distribution.  If these
IMBH-BH mergers were 2-body processes, we might expect the
eccentricity of the orbit to be very nearly circular right before the
black holes merge as gravitational radiation grinds away the orbital
angular momentum~\citep{Peters:1963ux}. However, few body encounters
are much more common within a globular cluster because the interaction
cross-section is much larger~\citep{Heggie:book}. Therefore, many BHs
are shepherded into mergers with an IMBH through exchange with lower
mass black holes~\citep{Miller:2002pi}, and the resulting eccentricity
can be quite large~\citep{Gultekin:2006tb}. In fact, simulations have
shown that rare interactions can yield mergers with $e>0.999$, and
such a highly eccentric orbit can become even {\emph{more}} eccentric
through gravitational radiation emission~\citep{Peters:1963ux,Kennefick:1998ab}.
Therefore, though rare, {\emph{highly eccentric binary black hole mergers can take
    place in astrophysically relevant systems}}. To assign
eccentricities to each merger, we use the simulation results
of~\citet{Gultekin:2006tb}, which empirically characterizes the
eccentricity distribution as a function of the mass ratio of the
encounter. Note, though, that equation~(\ref{eqn:Fit}) is really only
valid in the small eccentricity regime and, thus, it may be true that
the kicks can be even higher than those studied here for such nearly
radial orbits.

\section{IMBH Survival and Occupation Fraction}

In order to determine the probability that an IMBH survives the short
merger epoch after formation, our simulations consist of $10^6$ Monte
Carlo realizations of an N-step merger chain. The merger chains are
tailored to mimic the initial conditions and encounters
predicted by current IMBH formation theories within proto-globular and stellar cluster environments as described in the
previous section.

During a merger, gravitational waves radiate not only linear momentum,
but also angular momentum and energy or mass. Fully relativistic
numerical simulations suggest that $\sim 25\%$ of the angular momentum
(defined at the innermost stable circular orbit) can be radiated
during the merger
~\citep{Pretorius:2005gq,Campanelli05a,Baker:2006yw,Herrmann:2006ks}.
In addition, the mass of the merger product is only $\sim 95\%$ the
mass of the two progenitor black holes
~\citep{Pretorius:2005gq,Campanelli05a,Baker:2006yw,Herrmann:2006ks}.
Hence, after each step within a particular merger tree, we adjust the
total spin and mass of the merger remnant to account for these losses.

\begin{figure}
\epsfig{file=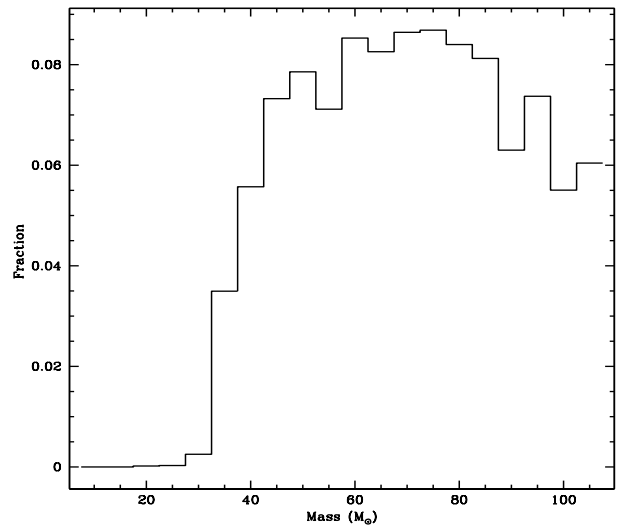, height=3.in, width=3.4in}
\caption{ A histogram of the black hole masses that eject a 
$1000 M_\odot$ IMBH. In this experiment, we assumed a Kroupa IMF, and
a uniform spin orientations and magnitude. Ejecting a $1000 M_\odot$ black hole
is possible with masses as low as $20 M_\odot$, but the overwhelming majority 
of ejections result from black holes with masses $>50 M_\odot$.}
\label{fig:ejectfrac}
\end{figure}

\begin{figure}
\epsfig{file=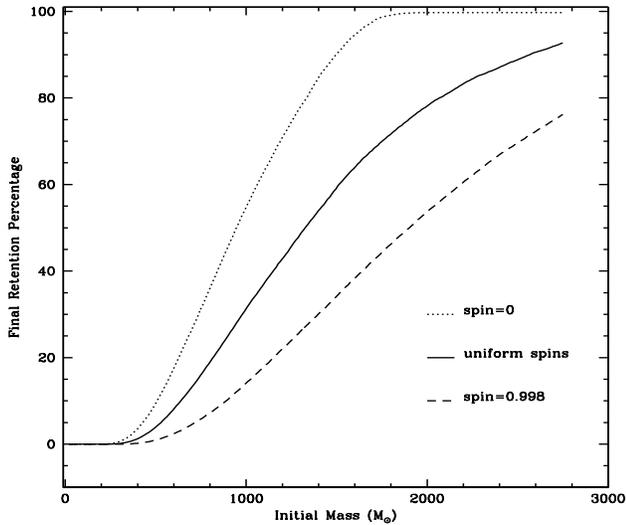, height=3.in, width=3.4in}
\caption{
  Percentage of black hole retention as a function of initial black
  hole mass within the Milky Way globular cluster system.  A black
  hole is defined as 'retained' if it survives 25 collisions with a BH
  selected from a Kroupa IMF with random spin orientations.
  The three lines on this
  figure represent different assumptions for the spins of the black
  hole.}
\label{fig:retain}
\end{figure}

\begin{figure}
\epsfig{file=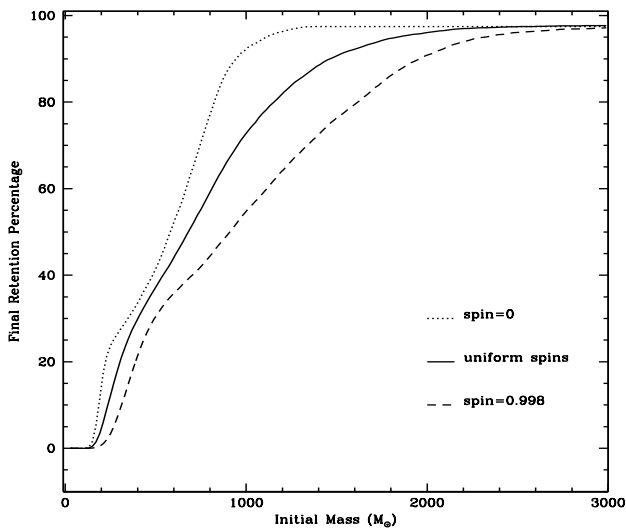, height=3.in, width=3.4in}
\caption{   Percentage of black hole retention as a function of initial black
  hole mass within the Milky Way globular cluster system. 
In this experiment, we assumed a black hole mass function 
described by model C1 of \citet{Belczynski:2006bh}, and
a uniform spin orientations and magnitude.}
\label{fig:retainbhmf}
\end{figure}

Figure~\ref{fig:retain} demonstrates the retention percentage after 25
mergers with black holes selected from a Kroupa IMF as a function of
the initial IMBH mass. This figure indicates that retaining an IMBH of
less than $1000 M_\odot$ occurs less than $33\%$ of the time for the
given distribution of black hole masses.  While the large number of
the lower-mass black holes dominate the total number of mergers, the
rare mergers with massive stellar-mass black holes dominate the
ejections -- and this trend only strengthens as the IMBH mass increases. 
 For example, figure~\ref{fig:ejectfrac} shows that 
for a $1000~\Msun$ seed, nearly all ejections come from black holes with 
mass $M > 30~\Msun$ and that most come from those with mass $M \sim 70~\Msun$.
Naturally, this implies that if $> 30 M_\odot$ black holes are extremely
rare in primordial globular clusters, the retention fraction increases 
dramatically~\citep{2001ApJ...554..548F}, making gravitational recoil 
ineffective in ejecting massive IMBHs. Figure~\ref{fig:retainbhmf} shows that it is easier to retain an 
IMBH of less than $1000 M_\odot$ with the shallower Belczynski black 
hole mass function, with only $30\%$ ejected.

Recall that each formation channel is expected to take place in a
particular proto-globular cluster environment ({\emph{e.g.},} single
runaways ought to form in very dense systems.) Since predictions of
pre-core collapse clusters indicate that the densities are $\sim 10$
times higher than today, we can use current observations of the
central luminosity density~\citep{Harris:96globs} to infer the initial
conditions of the globular cluster. Then, we can estimate on a
case-by-case basis what IMBH formation channel might have been
appropriate for each globular cluster. If we assume that {\it every}
globular cluster forms an IMBH with a mass that is consistent with its
presumed formation channel (see previous section), we can estimate the number of surviving
IMBHs within the Milky Way globular cluster system.  We find that
$\sim 24\%$ of the Milky Way globular clusters can host single
runaways, while $\sim 25\%$ may host multiple runaways. If, thereafter,
the IMBH must survive $25$ mergers selected from a Kroupa IMF, our
study indicates that no more than $5$ globular clusters have retained
an IMBH, while our Belczynski black hole mass function yields $<25$ IMBH-embedded
globulars.

Such conclusions can be split into the number that survive per
formation channel. Doing so, we learn that the only way to retain 
IMBHs during the short BH merger phase is with one rather massive 
initial seed, such as those produced
in a stellar runaway; any process that relies of the formation of
~$\ltsim 500 M_\odot$
black holes will be ineffective, as these are ejected from a globular cluster before forming an
IMBH. One way to lower the retention mass threshold
would be to increase the globular cluster escape velocity. We explore
this effect in figure~\ref{Fig-ret-prob}, where we observe that the
escape velocity must reach $\sim 80~\KMS$ to attain complete retention 
of $500 M_\odot$ black holes.  

A BH mass function
that is tipped more strongly to lower masses would also increase the
retention rate -- note that a Kroupa IMF with no mass loss yields an
average BH mass of $\sim 20 M_\odot$, while allowing for mass loss
will make high mass ratio mergers less likely.  
If there are no black holes $>20 M_\odot$ in clusters, then any IMBH seed larger than
$M \gtsim 600~M_\odot$ will remain.  This critical mass ratio is comparable 
to that necessary to avoid ejection from three-body dynamical kicks 
alone \citep{Gultekin:2004gi,Oleary:2005bm,Gultekin:2006tb}.
Since runaway stellar mergers preferentially include the cluster's
heaviest stars, which are the precursors to the heaviest black holes
in the cluster, the largest black holes used in our calculations may
not be present in the cluster.

\section{Conclusion}

Our studies indicate that it is a challenge to retain IMBHs during the
barrage of BH mergers expected in an early globular cluster. We find
no scenario that guarantees 100$\%$ retention for IMBHs up to 3000
$M_\odot$ when the interacting BHs are selected from a reasonable
distribution of spins, orientations, and mass ratios.  However, the
more massive the initial IMBH seed, the better the retention
probability. This may indicate that any IMBH observed in
globular clusters today would most likely have originated from an
early stellar runaway channel. 

The results obtained here might be used to test
some IMBH formation mechanisms. For example, \citet{Fregeau:06imbhgw} and \citet{Gurkan:2005mb} find that
with a binary fraction above 10\% a single cluster can host multiple
runaways, leading to multiple IMBHs.  The heaviest two such IMBHs will
find their way to the center of the cluster and merge very quickly
($\sim 1~{\rm Myr}$).
Because such IMBHs are nearly equal in mass and likely have high
spins, they will escape the cluster upon
merger.  This leaves the cluster without a seed unless a third runaway
of sufficient mass occurs.  The problem is that the third runaway is
often much less massive and therefore difficult to retain, as we have shown.

In fact, since lower mass IMBHs so easily escape globular clusters, 
if globular cluster observations find that they are {\it not} rare, it may be
possible to constrain the IMBH merger history, as well.
For example, if many low mass IMBHs exist within clusters, we may rule out 
low mass ratio mergers -- this would imply that there is no high mass tail in the BH IMF. 
Alternatively, the lower the spin, the better the retention as  
figure~\ref{fig:retain} shows; if low mass IMBHs are found in 
large numbers within clusters, and if BHs are found with $>20 M_\odot$,
we may have to explore ways in which the black holes spin down and align 
within a gas-poor globular.

Making some very simple assumptions for the primordial globular
cluster environment, such as the BH IMF and central density structure,
we have estimated that less than $5$ globular clusters retain their
IMBHs within the Milky Way -- even if every one hosted an initial IMBH seed. Naturally, there are many uncertainties folded into
this estimate, such as the shape of the primordial globular cluster
IMF, the degree of mass loss in low metallicity systems and its effect
on the BH IMF, and the detailed role that few-body/BH interactions
play in shaping early globular cluster structure. As more work is done
on these areas, we can revise our estimates using more
generic results. For example, if the BH IMF were instead narrowly
peaked around $10 M_\odot$~\citep{2001ApJ...554..548F} and if the seed black hole were a massive
supernova remnant, our calculations indicate that about $25\%$ of the
Milky Way globular clusters could retain black holes as small as $\sim 400 M_\odot$. 
An IMBH of a few hundred $M_\odot$ may not leave an
electromagnetically observable impact on the surrounding globular
cluster, as the dynamical effects on the surrounding stars may also be
produced by a high binary fraction \citep{Trenti:07imbh}. However,
when stellar mass compact objects merge with these smaller mass IMBHs,
they will produce a strong gravitational wave signals that should be
detectable with Advanced LIGO~\citep{mandel:07imri}.

With so many small black holes having been ejected from their host
globular clusters, we speculate that $\sim 100$ {\emph{rogue}}
black holes are swarming about in the Milky Way halo with masses from
$\sim 100 - 1000 M_\odot$, and with velocities mostly on the order of a 
few hundred $\KMS$. The number of rogues could be in the thousands if, as has been
suggested, the current globular cluster population is a small fraction
of the total number originally created \citep{Aguilar:88gc}. 

Although we have focused on IMBHs in the Galactic globular cluster
population, the same processes may occur in other galaxies.
Extragalactic ULXs, which may be powered by $\sim 10^{2}~M_\odot$
IMBHs, are frequently found near, but not in young stellar clusters
\citep[e.g.,][]{fabbianoetal01,2005ApJS..157...59L}.  Note, though, that the stellar clusters associated with ULXs are not always the dense stellar systems required by the IMBH formation models considered here \citep{liuetal07}.  If such IMBHs did form within the nearby
clusters, they may be ejected from gravitational wave kicks coming from
mergers with stellar-mass black holes, especially as they would merge
with the most massive black holes first.  While this would explain
their separation from the cluster center, it would not explain the fact that ULXs are
accreting sources; it is not clear how an IMBH would pick up a
companion on its way out of the cluster and is unlikely to retain a stellar companion close enough to overfill its Roche lobe.  
An IMBH with a stellar companion, however, may be ejected from the host
cluster through few-body Newtonian dynamical kicks that harden the binary until it begins accreting \citep{Gultekin:2004gi,Oleary:2005bm,Gultekin:2006tb,Blecha:06imbhulx}.

Even if the ejected IMBH is not accreting gas as a ULX, electromagnetic observations
may still detect rogue black holes. For instance, if the IMBH were to
carry a few massive stars along as it is ejected, our results indicate
a kinematically fast subpopulation of massive stars near globular
clusters.  The ejected black holes may leave a temporary imprint on
the globular cluster as well. Since they are ejected from the system
impulsively, it is likely that the globular cluster core would temporarily
expand.
Direct simulations remain to be done to determine how the globular
cluster responds to the ejection of an IMBH.

The consequences of these large recoil velocities may also affect SMBH
assembly. The most likely candidates for SMBH seeds are $\sim 10^3
M_\odot$ Pop~III stellar remnants at redshifts~$z \gtsim 12-20$
\citep{Heger:03sn,Volonteri:03smbh,Islam02,Wise:05snpop3,Micic:2007sb}.
These relic seeds are predicted to form at the centers of low mass
dark matter halos ($\sim 4 \times 10^6 M_\odot$).  As dark matter
halos hierarchically merge to assemble the galaxy, the seed black
holes sink to the center through dynamical friction and eventually
merge. With kick velocities in the range of $\sim 10^{2}-10^{3} \KMS$, it may
also be difficult to retain seed SMBHs in high redshift low mass dark
matter halos. We plan to explore black hole retention and possible
kick suppression mechanisms at the low mass end of the halo mass
function using high resolution cosmological N-body simulations in our
next paper.

\acknowledgments
KHB and KG wish to thank M. Coleman Miller for several very 
helpful discussions.
This work was supported by
NSF grant PHY-0354821 to Deirdre Shoemaker. NY is supported by 
NSF grant PHY-0555628 to Ben Owen.

\bibliography{references}

\end{document}